%% file: DESY-04-056.tex
\documentclass[zpreprint,zbstdefault]{zeus_paper}
\usepackage[english]{babel}

\input zeus_def.tex
\newcommand{\ks}{K^{0}_{S}}
\newcommand{\ksp}{K^{0}_{S}\> p}
\newcommand{\kspb}{K^{0}_{S}\> \bar{p}}
\newcommand{\ksppb}{K^{0}_{S}\> p\>(\bar{p})}

\input  DESY-04-056-cit.tex
\begin{document}
%------------------------------------------------------------------------------
%       Title sheet
%------------------------------------------------------------------------------
\include{DESY-04-056-tit}
%------------------------------------------------------------------------------
%       Authors
%------------------------------------------------------------------------------
\include{auth119_out}
%------------------------------------------------------------------------------
%       Text
%------------------------------------------------------------------------------
\include{DESY-04-056-txt}
%------------------------------------------------------------------------------
%       Bibliography
%------------------------------------------------------------------------------
\include{DESY-04-056-ref}
%------------------------------------------------------------------------------
%       Tables
%------------------------------------------------------------------------------
\include{DESY-04-056-tab}
%------------------------------------------------------------------------------
%       Figures
%------------------------------------------------------------------------------
\include{DESY-04-056-fig}

%
%       ... that's it
%
\end{document}

%% file: zeus_def.tex
\newcommand{\ZcoosysB}{%
The ZEUS coordinate system is a right-handed Cartesian system, with the $Z$
axis pointing in the proton beam direction, referred to as the ``forward
direction'', and the $X$ axis pointing left towards the centre of HERA.
The coordinate origin is at the nominal interaction point.\xspace}
\newcommand{\Zpsrap}{}

\newcommand{\ZcoosysfnBeta}{\footnote{\ZcoosysB\Zpsrap}}
\newcommand{\Zdetdesc}{%
A detailed description of the ZEUS detector can be found 
elsewhere~\cite{zeus:1993:bluebook}. A brief outline of the 
components that are most relevant for this analysis is given
below.\xspace}
\newcommand{\Zctddesc}[1]{%
Charged particles are tracked in the central tracking detector (CTD)~\citeCTD,
which operates in a magnetic field of $1.43\Tesla$ provided by a thin 
superconducting solenoid. The CTD consists of 72~cylindrical drift chamber 
layers, organised in nine superlayers covering the polar-angle#1 region 
\mbox{$15^\circ<\theta<164^\circ$}. The transverse-momentum resolution for
full-length tracks is $\sigma(p_T)/p_T=0.0058p_T\oplus0.0065\oplus0.0014/p_T$,
with $p_T$ in $\Gev$.}
\newcommand{\Zcaldesc}{%
The high-resolution uranium--scintillator calorimeter (CAL)~\citeCAL consists 
of three parts: the forward, the barrel and the rear 
calorimeters. 
The smallest subdivision of
the calorimeter is called a cell.  The CAL energy resolutions, as measured under
test-beam conditions, are $\sigma(E)/E=0.18/\sqrt{E}$ for electrons and
$\sigma(E)/E=0.35/\sqrt{E}$ for hadrons, with $E$ in $\Gev$.}
\chardef\usc=95
\chardef\til=126
\catcode`\@=11 % @ signs are now treated as letters
\DeclareRobustCommand\xdotspace{\futurelet\@let@token\@xdotspace}
\def\@xdotspace{%
  \ifx\@let@token.\else
  \ifx\@let@token\bgroup.\else
  \ifx\@let@token\egroup.\else
  \ifx\@let@token\/.\else
  \ifx\@let@token\ .\else
  \ifx\@let@token~.\else
  \ifx\@let@token!.\else
  \ifx\@let@token,.\else
  \ifx\@let@token:.\else
  \ifx\@let@token;.\else
  \ifx\@let@token?.\else
  \ifx\@let@token/.\else
  \ifx\@let@token'.\else
  \ifx\@let@token).\else
  \ifx\@let@token-.\else
  \ifx\@let@token\@xobeysp.\else
  \ifx\@let@token\space.\else
  \ifx\@let@token\@sptoken.\else
   .\space
   \fi\fi\fi\fi\fi\fi\fi\fi\fi\fi\fi\fi\fi\fi\fi\fi\fi\fi}
\catcode`\@=12 % @ signs are no longer letters

\newcommand{\stru}[2]{%
   \relax\ifmmode\hbox{\vrule height#1 depth#2 width0pt}%
   \else\vrule height#1 depth#2 width0pt\fi}
\newcommand{\Ronum}[1]{\uppercase\expandafter{\romannumeral#1}}
\newcommand{\ronum}[1]{\expandafter{\romannumeral#1}}
\DeclareRobustCommand{\LaTeXZ}{%
  \LaTeX\kern-.05em4\kern-.1em
  {\raisebox{-0.2ex}{$\scriptstyle\text{ZEUS}$}}\xspace}
\DeclareMathAlphabet{\mathbf}{OT1}{cmr}{bx}{sl}
\newcommand{\eVdist}{\kern-0.06667em}

\newcommand{\Gev}{{\text{Ge}\eVdist\text{V\/}}}

\newcommand{\mev}{{\,\text{Me}\eVdist\text{V\/}}}
\newcommand{\gev}{{\,\text{Ge}\eVdist\text{V\/}}}

\newcommand{\Tesla}{\,\text{T}}

\newcommand{\slashfrac}[2]{%
  \raisebox{0.5ex}{\ensuremath #1}\kern-0.12em/\kern-0.08em
  \raisebox{-.8ex}{\ensuremath #2}}
\newcommand{\sqr}[3]{%
    {\vcenter{\hrule height.#3ex\hbox{\vrule width.#2ex height#1ex
     \kern#1ex\vrule width.#3ex}\hrule height.#2ex}}}

\catcode`\@=11 % @ signs are now treated as letters
\newcommand{\parenbar}{\mathpalette\p@renb@r}
\def\p@renb@r#1#2{\vbox{%
  \ifx#1\scriptscriptstyle \dimen@.7em\dimen@ii.2em\else
  \ifx#1\scriptstyle \dimen@.8em\dimen@ii.25em\else
  \dimen@1em\dimen@ii.4em\fi\fi \offinterlineskip
  \ialign{\hfill##\hfill\cr
    \vbox{\hrule width\dimen@ii}\cr
    \noalign{\vskip-.3ex}%
    \hbox to\dimen@{$\mathchar300\hfil\mathchar301$}\cr
    \noalign{\vskip-.3ex}%
    $#1#2$\cr}}}
\catcode`\@=12 % @ signs are no longer letters

\newcommand{\IP}{{\rm I$\kern-0.01667em$P}\xspace}

\mathchardef\qsm=63
\mathchardef\pls=43
\mathchardef\mns=512
\mathchardef\plm=518
\mathchardef\eql=61
\mathchardef\smallleft=300
\mathchardef\smallright=301
\mathchardef\les=316
\mathchardef\gre=318
\mathchardef\leq=532
\mathchardef\grq=533
\catcode`\@=11 % @ signs are now treated as letters
\newcounter{pict@width}
\newcounter{pict@height}
\newlength{\pict@scale}
\newcommand{\psfigadd}[4]{%
\setcounter{pict@width}{1*\ratio{#2+\pict@scale/2}{\pict@scale}}
\setcounter{pict@height}{1*\ratio{#3+\pict@scale/2}{\pict@scale}}
\setlength{\unitlength}{\pict@scale}
\hbox to #2{\hspace{-\fill}\begin{picture}(\thepict@width,\thepict@height)
\put(0,0){\psfig{figure=#1,width=#2,height=#3,clip=}}
\SetScale{0.283466457}
\SetWidth{1.763889}
{#4}
\end{picture}}
}
\newcounter{pict@widthfst}
\newcounter{pict@widthscd}
\newcounter{pict@widthtot}
\newcommand{\psfigaddtwo}[7]{%
\setcounter{pict@widthfst}{1*\ratio{#2+\pict@scale/2}{\pict@scale}}
\setcounter{pict@widthscd}{1*\ratio{#2+#4+\pict@scale/2}{\pict@scale}}
\setcounter{pict@widthtot}{1*\ratio{#2+#4+#6+\pict@scale/2}{\pict@scale}}
\setcounter{pict@height}{1*\ratio{#3+\pict@scale/2}{\pict@scale}}
\setlength{\unitlength}{\pict@scale}
\hbox{\hspace{-\fill}\begin{picture}(\thepict@widthtot,\thepict@height)
\put(0,0){\psfig{figure=#1,width=#2,height=#3,clip=}}
\put(\thepict@widthscd,0){\psfig{figure=#5,width=#6,height=#3,clip=}}
\SetScale{0.283466457}
\SetWidth{1.763889}
{#7}
\end{picture}}
}
\newcommand{\psfigror}[4]{%
\setcounter{pict@width}{1*\ratio{#2+\pict@scale/2}{\pict@scale}}
\setcounter{pict@height}{1*\ratio{#3+\pict@scale/2}{\pict@scale}}
\setlength{\unitlength}{\pict@scale}
\hbox{\begin{picture}(\thepict@width,\thepict@height)
\put(0,\thepict@height){\psfig{figure=#1,width=#3,height=#2,clip=,angle=270}}
\SetScale{0.283466457}
\SetWidth{1.763889}
{#4}
\end{picture}}
}
\newcommand{\psfigrol}[4]{%
\setcounter{pict@width}{1*\ratio{#2+\pict@scale/2}{\pict@scale}}
\setcounter{pict@height}{1*\ratio{#3+\pict@scale/2}{\pict@scale}}
\setlength{\unitlength}{\pict@scale}
\hbox{\begin{picture}(\thepict@width,\thepict@height)
\put(0,0){\psfig{figure=#1,width=#3,height=#2,clip=,angle=90}}
\SetScale{0.283466457}
\SetWidth{1.763889}
{#4}
\end{picture}}
}
\catcode`\@=12 % @ signs are no longer letters
\newlength\listtextwidth

\catcode`\@=11 % @ signs are now treated as letters
\newlength{\@tabfninsert}
\newlength{\@tabfnwidth}
\newcommand{\tabfootnote}[2]{%
  \setlength{\@tabfninsert}{0.8em}
  \setlength{\@tabfnwidth}{\textwidth}
  \addtolength{\@tabfnwidth}{-\@tabfninsert}
  \addtolength{\@tabfnwidth}{-0.4em}
  \noindent\makebox[\@tabfninsert][r]{\footnotesize$^{#1}$\hfil}\hfill%
  \parbox[t]{\@tabfnwidth}{\footnotesize #2\hfill}}
\catcode`\@=12 % @ signs are no longer letters

%% file: DESY-04-056-cit.tex
\def\citeCTD{{\cite{%
nim:a279:290,*npps:b32:181,*nim:a338:254%
}}\xspace}
\def\citeCAL{{\cite{%
nim:a309:77,*nim:a309:101,*nim:a321:356,*nim:a336:23%
}}\xspace}

%% file: DESY-04-056-tit.tex
\prepnum{DESY--04--056} 

\title{
Evidence for a narrow baryonic state decaying to $\ksp$ and $\kspb$
in deep inelastic scattering at HERA }
                    
\author{ZEUS Collaboration}

\date{March 27, 2004}

\abstract{

A resonance search has been made in the $\ksp$ and $\kspb$
invariant-mass spectrum measured with the ZEUS detector at HERA using
an integrated luminosity of 121 pb$^{-1}$. The search was performed in
the central rapidity region of inclusive deep inelastic scattering at
an $ep$ centre-of-mass energy of 300--318 GeV for exchanged photon
virtuality, $Q^2$, above 1 $\gev^2$.  Recent results from fixed-target
experiments give evidence for a narrow baryon resonance decaying to
$K^+ n$ and $\ksp$, interpreted as a pentaquark. The results
presented here support the existence of such state, with a mass of
$1521.5\pm 1.5 ({\rm stat.})^{+2.8}_{-1.7} ({\rm syst.}) \mev$ and a
Gaussian width consistent with the experimental resolution of
$2\mev$. The signal is visible at high $Q^2$ and, for $Q^2 >
20\gev^2$, contains $221 \pm 48$ events. The probability of a similar
signal anywhere in the range 1500--1560 MeV arising from fluctuations
of the background is below $6 \times 10^{-5}$.  }

\makezeustitle

%% file: auth119_out.tex
\pagenumbering{Roman}                                                                              
                                    % this "%"s are for cosmetics only                             
                                                   %                                               
\begin{center}                                                                                     
{                      \Large  The ZEUS Collaboration              }                               
\end{center}                                                                                       
  S.~Chekanov,                                                                                     
  M.~Derrick,                                                                                      
  J.H.~Loizides$^{   1}$,                                                                          
  S.~Magill,                                                                                       
  S.~Miglioranzi$^{   1}$,                                                                         
  B.~Musgrave,                                                                                     
  J.~Repond,                                                                                       
  R.~Yoshida\\                                                                                     
 {\it Argonne National Laboratory, Argonne, Illinois 60439-4815}, USA~$^{n}$                       
\par \filbreak                                                                                     
  M.C.K.~Mattingly \\                                                                              
 {\it Andrews University, Berrien Springs, Michigan 49104-0380}, USA                               
\par \filbreak                                                                                     
  N.~Pavel \\                                                                                      
  {\it Institut f\"ur Physik der Humboldt-Universit\"at zu Berlin,                                 
           Berlin, Germany}
% ...
\newpage                                                                                           
$^{\    1}$ also affiliated with University College London, London, UK \\                          
% ...
                                                           %                                       
                                                           %                                       
% \par         % if index listing & table fit to 1 page, put gap here                              
\newpage   % alternatively: go to newpage, if page is too small                                    
                                                           %                                       
% \institute_references_start    % do not touch or move this line !                                
                                                           %                                       
\begin{tabular}[h]{rp{14cm}}                                                                       
$^{a}$ &  supported by the Natural Sciences and Engineering Research                               
          Council of Canada (NSERC) \\                                                             
$^{b}$ &  supported by the German Federal Ministry for Education and                               
          Research (BMBF), under contract numbers HZ1GUA 2, HZ1GUB 0, HZ1PDA 5, HZ1VFA 5\\         
$^{c}$ &  supported by the MINERVA Gesellschaft f\"ur Forschung GmbH, the                          
          Israel Science Foundation, the U.S.-Israel Binational Science                            
          Foundation and the Benozyio Center                                                       
          for High Energy Physics\\                                                                
$^{d}$ &  supported by the German-Israeli Foundation and the Israel Science                        
          Foundation\\                                                                             
$^{e}$ &  supported by the Italian National Institute for Nuclear Physics (INFN) \\                
$^{f}$ &  supported by the Japanese Ministry of Education, Culture,                                
          Sports, Science and Technology (MEXT) and its grants for                                 
          Scientific Research\\                                                                    
$^{g}$ &  supported by the Korean Ministry of Education and Korea Science                          
          and Engineering Foundation\\                                                             
$^{h}$ &  supported by the Netherlands Foundation for Research on Matter (FOM)\\                   
$^{i}$ &  supported by the Polish State Committee for Scientific Research,                         
          grant no. 620/E-77/SPB/DESY/P-03/DZ 117/2003-2005\\                                      
$^{j}$ &  partially supported by the German Federal Ministry for Education                         
          and Research (BMBF)\\                                                                    
$^{k}$ &  supported by RF President grant N 1685.2003.2 for the leading                            
          scientific schools and by the Russian Ministry of Industry, Science                      
          and Technology through its grant for Scientific Research on High                         
          Energy Physics\\                                                                         
$^{l}$ &  supported by the Spanish Ministry of Education and Science                               
          through funds provided by CICYT\\                                                        
$^{m}$ &  supported by the Particle Physics and Astronomy Research Council, UK\\                   
$^{n}$ &  supported by the US Department of Energy\\                                               
$^{o}$ &  supported by the US National Science Foundation\\                                        
$^{p}$ &  supported by the Polish Ministry of Scientific Research and Information                  
          Technology, grant no. 112/E-356/SPUB/DESY/P-03/DZ 116/2003-2005\\                        
$^{q}$ &  supported by the Polish State Committee for Scientific Research,                         
          grant no. 115/E-343/SPUB-M/DESY/P-03/DZ 121/2001-2002, 2 P03B 07022\\                    
\end{tabular}                                                                                      
                                                           %                                       
% \institute_references_end     % do not touch or move this line !                                 

%% file: DESY-04-056-txt.tex
\pagenumbering{arabic} 
\pagestyle{plain}
% ----------------------------------------------------------------------------
%       Introduction
% ----------------------------------------------------------------------------
\section{Introduction}
\label{sec-int}

Recent results from fixed-target experiments give evidence for the
existence of a narrow baryon resonance with a mass of approximately
1530 MeV and positive strangeness
\cite{prl91012003,*plb572:127,*prl91kub,*prl92:032001}, 
seen in the $K^+ n$ decay channel.
%*hep-ex0307088,
These results have triggered new interest in baryon spectroscopy since
this baryon is manifestly exotic; it cannot be composed of three
quarks, but may be explained as a bound state of five quarks, i.e. as
a pentaquark, $\Theta^+ = uudd\bar{s}$. A narrow baryonic resonance
close to the observed mass is predicted in the chiral soliton
model\cite{zp:a359:305}. The quantum numbers of this state also permit
decays to $\ksp$ and $\kspb$ (denoted as $\ksppb$). Evidence for a
corresponding signal has been seen
\cite{pan66:1715,*hep-ex-0309042,*hep-ex-0401024,*plb585:213,*hep-ex-0403011} in this
channel by other experiments.  Evidence for two other pentaquark
states has also been reported recently
\cite{prl92:042003,hep-ex-0403017}.

The Particle Data Group (PDG)\cite{pr:d66:010001} lists a number of
`$\Sigma$ bumps', unestablished resonances observed with low
significance by previous fixed-target experiments.
The possible presence of these resonances in the mass region close to
the production threshold of the $\ksppb$ final state complicates the
search for pentaquarks in this decay channel.

The $\Theta^+$ state and the $\Sigma$ bumps discussed above have never
been observed in high-energy experiments, where hadron production is
dominated by fragmentation. 
This paper presents the results of a search for narrow states in the
$\ksppb$ decay channel in the central rapidity region of high-energy
$ep$ collisions,
where particle production is not expected to be influenced
by the baryon number in the initial state. 
The analysis was performed using deep inelastic
scattering (DIS) events measured with exchanged-photon virtuality
$Q^2\ge 1\gev^2$.  The data sample, collected with the ZEUS detector
at HERA, corresponds to an integrated luminosity of 121 pb$^{-1}$,
taken between 1996 and 2000. This sample is the sum of 38 pb$^{-1}$ of
$e^+p$ data taken at a centre-of-mass energy of $300 \gev$ and 68
pb$^{-1}$ taken at $318
\gev$, plus 16 pb$^{-1}$ of $e^-p$ data taken at $318 \gev$.

%%%%%%%%%%%%%%%%%%%%%%%%%%%%%%%%%%%%%
%       Experimental set-up
%%%%%%%%%%%%%%%%%%%%%%%%%%%%%%%%%%%%%
\section{Experimental set-up}
\label{sec-exp}

\Zdetdesc

\Zctddesc\ZcoosysfnBeta\
To estimate the ionization energy loss per unit length, $dE/dx$, of
particles in the CTD\cite{pl:b481:213,*epj:c18:625}, the truncated
mean of the anode-wire pulse heights was calculated, which removes the
lowest $10\%$ and at least the highest $30\%$ depending on the number
of saturated hits. The measured $dE/dx$ values were corrected by
normalising to the average $dE/dx$ for tracks around the region of
minimum ionisation for pions, $0.3~<~p~<~0.4$~GeV. Henceforth, $dE/dx$
is quoted in units of minimum ionising particles (mips).  The $dE/dx$
distribution for electrons has a roughly Gaussian shape centred about
$dE/dx~\sim~1.4$~mips with width 0.14~mips, corresponding to a
resolution of $\sim10\%$.

\Zcaldesc
$\>$ A presampler \cite{nim:a382:419,*magill:bpre} mounted in front of
the calorimeter was used to correct the energy of the scattered
electron\footnote{Henceforth the term
electron is used to refer both to electrons and positrons.}. 
The position of electrons scattered
close to the electron beam direction is determined by a
scintillator-strip detector
\cite{nim:a401:63}.

The luminosity was measured using the bremsstrahlung process $ep \to e
p \gamma$ with the luminosity
monitor~\cite{Desy-92-066,*zfp:c63:391,*acpp:b32:2025}, a
lead--scintillator calorimeter placed in the HERA tunnel at $Z = -107$
m.

%%%%%%%%%%%%%%%%%%%%%%%%%%%%%%%%%%%%%%%%%%%%%%%%%%%%%%
\section{Event simulation}
\label{sec:evsim}
%%%%%%%%%%%%%%%%%%%%%%%%%%%%%%%%%%%%%%%%%%%%%%%%%%%%%%

Inclusive DIS events were generated using the {\sc Ariadne} 4.08 Monte
Carlo (MC) model \cite{cpc:71:15} interfaced with {\sc Heracles} 4.5.2
\cite{cpc:69:155-tmp-3cfb28c9} via the {\sc Djangoh} 1.1 program \cite{spi:www:djangoh11}
in order to incorporate first-order electroweak corrections.  The {\sc
Ariadne} program uses the Lund string model
\cite{prep:97:31} for hadronisation, as implemented in 
{\sc Jetset} 7.4\cite{cpc:82:74}.

Before detector simulation, the $\ksppb$ invariant-mass distribution
was calculated from the true $\ks$ and (anti)protons in the mass range
up to 1700 MeV. No peaks were found, indicating that no reflection
from known decays are expected to generate a narrow peak in these
decay channels.

The generated events were passed through a full simulation of the
detector using {\sc Geant} 3.13
\cite{tech:cern-dd-ee-84-1} and processed with the same reconstruction
program as used for the data.  The detector-level MC samples were then
selected in the same way as the data. The generated MC statistics were
about three times higher than those of the data.

%%%%%%%%%%%%%%%%%%%%%%%%%%%%%%%%%%%%%%%%%%%%%%%%%%%%%%%%%%%%%%%%%%%%
\section{Event sample}
\label{sec:data}
%%%%%%%%%%%%%%%%%%%%%%%%%%%%%%%%%%%%%%%%%%%%%%%%%%%%%%%%%%%%%%%%%%%

The search was performed using DIS events with $Q^2\ge 1\gev^2$, the
largest event sample for which no explicit trigger requirement
was imposed on the hadronic final state.  A three-level trigger
\cite{zeus:1993:bluebook} was used to select events online.  
At the third level, an electron with an energy greater than $4\gev$
and a position outside a box of $24\times 12$ cm$^2$ on the face of
the calorimeter was required. The trigger has a high acceptance for
$Q^2 \gtrsim 2$~GeV$^2$. However, data below $Q^2 \approx 20
\gev^2$ are strongly affected by prescales which were applied to the 
inclusive triggers to control data rates.

The Bjorken scaling variables $x$ and $y$, as well as $Q^2$, were
reconstructed using the electron method (denoted by the subscript
$e$), which uses measurements of the energy and angle of the scattered
electron, or using the Jacquet-Blondel (JB)
method\cite{proc:epfacility:1979:391}.  The scattered-electron
candidate was identified from the pattern of energy deposits in the
CAL \cite{nim:a365:508}.

The following requirements were imposed:

\begin{itemize}

\item[$\bullet$]
$Q_e^2\ge 1\gev^2$;

\item[$\bullet$]
$E_{e^{'}}\geq 8.5$ GeV, where $E_{e^{'}}$ is the
corrected energy of the scattered electron measured in the CAL;

\item[$\bullet$]
$35\> \leq\> \delta \>\leq\> 60$ GeV, where
$\delta=\sum E_i(1-\cos\theta_i)$,
$E_i$ is the energy of the $i$th calorimeter
cell, $\theta_i$ is its polar angle
and the sum runs over all cells;

\item[$\bullet$]
$y_{e}\> \leq\> 0.95$ and $y_{\mathrm{JB}}\> \geq\> 0.01$;

\item[$\bullet$]
$\mid Z_{\mathrm{vertex}} \mid  \le  50$ cm, where
$Z_{\mathrm{vertex}}$ is the vertex position
determined from the
tracks.
\end{itemize}

The present analysis is based on charged tracks measured in the
CTD. The tracks were required to pass through at least five CTD
superlayers and to have transverse momenta $p_T\ge 0.15\gev$ and
pseudorapidity in the laboratory frame $|\eta|\le 1.75$, restricting
the study to a region where the CTD track acceptance and resolution
are high.  Candidates for long-lived neutral strange hadron 
decaying to two charged
particles are identified by selecting pairs of oppositely charged
tracks, fitted to a displaced secondary vertex.
Events were required to have at least one such candidate.  

After these selection cuts, a sample of 1\,600\,000 events remained.

%%%%%%%%%%%%%%%%%%%%%%%%%%%%%%%%%%%%%%%%%%%%%%%%
\section{Reconstruction of $\ks$ candidates
\label{reco-ks}}
%%%%%%%%%%%%%%%%%%%%%%%%%%%%%%%%%%%%%%%%%%%%%%%%

The $\ks$ mesons were identified by their charged-decay mode,
$\ks\to\pi^{+}\pi^{-}$.  Both tracks were assigned the mass of the
charged pion and the invariant mass, $M(\pi^+\pi^-)$, of each track
pair was calculated. The $\ks$ candidates were selected by imposing the
following requirements:

\begin{itemize}
\item 
$M(e^{+}e^{-})\ge 50\mev$, where the electron mass was assigned
to each track, to eliminate tracks from photon conversions;
\item 
$M(p\pi)\ge 1121\mev$, where the proton mass was assigned to the track
  with higher momentum, to eliminate $\Lambda$ and
  $\bar{\Lambda}$ contamination of the $\ks$ signal;
\item 
$483\le M(\pi^+\pi^-)\le 513\mev$;
\item 
$p_T (\ks )\ge 0.3\gev$ and $|\eta (\ks ) |\le 1.5$. 
\end{itemize}

Figure~\ref{pap0} shows the invariant-mass distribution for $\ks$
candidates for $Q^2\ge 1\gev^2$.  A fit using two Gaussian functions
plus a first-order polynomial function was used.  The number of $\ks$
candidates was $866800 \pm 1000$, with a background under the peak
constituting approximately 6\% of the total number of candidates.  The
peak position was $m_{\ks}=498.12\pm0.01(\mathrm{stat.}) \mev$, which agrees
with the PDG value of $497.67\pm 0.03$\cite{pr:d66:010001} within
the calibration uncertainty of the CTD. The width is consistent with
the detector resolution.

%%%%%%%%%%%%%%%%%%%%%%%%%%%%%%%%%%%%%%%%%%%%%%%%
\section{Selection of $p(\bar{p})$ candidates
\label{reco-p}}
%%%%%%%%%%%%%%%%%%%%%%%%%%%%%%%%%%%%%%%%%%%%%%%%

The (anti)proton candidate selection used the energy-loss measurement
in the CTD, $dE/dx$.  Figure~\ref{pap0dedx} shows the $dE/dx$
distribution as a function of the track momentum for positive and
negative tracks.  Tracks fitted to the primary vertex were used, with
the exception of the scattered-electron track.  Tracks were selected
as described in Section~\ref{sec:data}.  In addition, only tracks with
more than $40$ CTD hits were used to ensure a good $dE/dx$
measurement.  The tracks were then selected by requiring $f\le
dE/dx\le F$, where $f$ and $F$, motivated by the Bethe-Bloch equation,
are the functions: $f=0.35 /p^{2}+0.8$, $F=1.0/p^{2}+1.2$ (for
positive tracks) and $f=0.3 /p^{2}+0.8$, $F=0.75 /p^{2}+1.2$ (for
negative tracks), where $p$ is the total track momentum in GeV.  These
cuts were found from an examination of $dE/dx$ as a function of
$p$ and were checked by studying (anti)proton candidate tracks from
$\Lambda (\bar{\Lambda})$ decays. The proton band was found to be
broader than that of the antiproton.  There is also a clear deuteron
band for positive tracks, which suggests a small contribution from
secondary interactions. To remove the region where
the proton band completely overlaps the pion band, the proton momentum
was required to be less than $1.5\gev$. Finally, a cut requiring
$dE/dx\ge 1.15$~mips was applied.

After these cuts, the purity of the proton sample, estimated from the
MC simulation, is around 60\%, varying from 96\% at low momentum to
17\% at the highest accepted momenta. Applying a higher $dE/dx$ cut
leads to higher purity, but reduces the acceptance for protons in the
high-momentum region and reduces the statistics.

%%%%%%%%%%%%%%%%%%%%%%%%%%%%%%%%%%%%%%%%%%%%%%%%%%%%%%%%%%%%
\section{Reconstruction of $\ksppb$ invariant mass 
\label{reco-m}}
%%%%%%%%%%%%%%%%%%%%%%%%%%%%%%%%%%%%%%%%%%%%%%%%%%%%%%%%%%%%

The $\ksppb$ invariant mass was obtained by combining $\ks$ and
(anti)proton candidates selected as described above, and fixing the
$\ks$ mass to the PDG value. The resolution of the $\ksppb$
invariant-mass measurement was estimated using MC simulations to be
$2.0\pm 0.5\mev$ in the region near $1530\mev$, for both the
$\ksp$ and the $\kspb$ channels.

The resolution was independently verified from the data by
reconstructing the $K^*(892)$ peak, assuming that the momentum and
angle resolution for (anti)protons is similar to that for pions.
Tracks passing the same angle and momentum cuts as the proton
candidate sample were taken from the pion band, assigned the pion
mass, and combined with the $\ks$ candidates.  A fit was performed
using a Breit-Wigner function convoluted with a Gaussian to describe
the resolution.  The width of the Breit-Wigner function was fixed to
the natural width of the $K^*$ resonance of $50.8\mev$  
\cite{pr:d66:010001}.  The resulting width of the Gaussian was
$5\pm 2(\mathrm{stat.})\mev$, independent of the pion charge. 

%%%%%%%%%%%%%%%%%%%%%%%%%%%%%%%%%%%%%%%%%%%%%%%%%%%
\section{Results}
%%%%%%%%%%%%%%%%%%%%%%%%%%%%%%%%%%%%%%%%%%%%%%%%%%%

The $\ksppb$ mass spectrum, in the range from threshold to 1700 MeV,
is shown in Fig.~\ref{pq} for various regions of the DIS phase space.
Figures~\ref{pq}a-\ref{pq}d show the mass distribution integrated
above a minimum $Q^2$ ranging from 1~GeV$^2$ to $50 \gev^2$.  The data
show signs of structure below about $1600\mev$.  For $Q^2 > 10
\gev^2$, a peak is seen in the mass distribution around 1520 MeV.
Figures
\ref{pq}e and \ref{pq}f show the $Q^2 > 1 \gev^2$ sample divided into 
two bins of the photon-proton centre-of-mass energy, $W$. A peak is
seen in the lower $W$ bin.

The expectation from the MC simulation, scaled to agree with the data
in the mass region above 1650 MeV, is also shown. If normalised to the
luminosity, the simulation lies below the data by a factor of
approximately two (not shown).  Even after scaling, the data are not
well described by the simulation for masses below 1600 MeV, and no
structure is seen in the simulated data. The PDG reports a $\Sigma$
bump at $1480
\mev$, and several states above $1550\mev$. None of these states are 
included in the simulation, which includes only well established
resonances.

Several functional forms were fit to the data for $Q^2 > 20\gev^2$
over the mass range from threshold up to 1700 MeV to estimate the
significance of the peak, as well as its position and width. A
background of the form
\begin{equation*}
P_1 (M- m_p - m_{\ks})^{P_2} \times (1 +P_3(M- m_p - m_{\ks})), 
\end{equation*}
where $M$ is the $\ksppb$ candidate mass, $m_p$ and $m_{\ks}$ are the
masses of the proton and the $\ks$, respectively, and $P_1$, $P_2$ and
$P_3$ are parameters, was found to give a good description of the data
when combined with Gaussians to describe the signal near 1520 MeV as
well as possible contributions from $\Sigma$ bumps. The parameters of
the Gaussians and the background were left free.  A bin-by-bin
$\chi^2$ minimisation was used. The results of the fit using the
background function plus one and two Gaussians are shown in
Table~\ref{table1}. Reducing the number of parameters in the
background function significantly reduces the quality of the fit.
Adding a third Gaussian does not significantly improve the fit.  The
result of the fit using two Gaussians is shown in Fig.~\ref{pqbar}.
The second Gaussian significantly improves the fit in the low mass
region, and has a mass of $1465.1\pm 2.9 (\mathrm{stat.})\mev$ and a
width of $15.5\pm 3.4 \mathrm{(stat.)}\mev$ and may correspond to the
$\Sigma (1480)$. However, the parameters and significance of any state
in this region are difficult to estimate due to the steeply falling
background close to threshold. The signal peak position is $1521.5 \pm
1.5 (\mathrm{stat.}) \mev$, with a measured Gaussian width 
$6.1\pm 1.6 (\mathrm{stat.})\mev$, 
consistent with the resolution.  The fit gives $221\pm
48$ events above the background, corresponding to $4.6\sigma$. The
equivalent estimate for the single Gaussian fit is $3.9\sigma$.  The
$\chi^2$ per degree of freedom for the fit over the whole fitted range
of masses is 35/44. Over the same range, the $\chi^2$ per degree of
freedom for the fit with no Gaussians is 69/50, which is an acceptable
fit. However, this value is dominated by contributions from the
high-mass region. A number of MC experiments were carried out, using
the background fit with zero gaussians as a starting distribution and
generating random data using Poisson statistics. The probability of a
fluctuation leading to a signal with $3.9\sigma$ significance or more
in the mass range 1500--1560 MeV and with a Gaussian width in the
range 1.5--12 MeV, was found to be $6
\times 10^{-5}$. The same exercise was carried out using the
background plus the 1465 MeV Gaussian as the starting distribution,
and the probability was found to be about a factor of ten lower.

The Gaussian function was replaced by a Breit-Wigner function
convoluted with a Gaussian to describe the peak near $1522\mev$.  The
width of the Gaussian distribution was fixed to the experimental
resolution to obtain an estimate of the intrinsic width of the signal.
The extracted width was $\Gamma=8\pm 4 (\mathrm{stat.})\mev$.

The fit was repeated for different values of the minimum $Q^2$ cut.
Above $Q^2 \approx 10\gev^2$ both the signal and background are
consistent with having a $1/Q^4$ dependence, similar to the inclusive
cross section. A MC study in which $\Sigma^+$ particles were
modified to have a mass of 1530 MeV and artificially forced to decay
to $\ksppb$ indicated that, at low $Q^2$, the impact of detector
acceptance on the visible cross section is important; for $Q^2 < 10
\gev^2$ the number of selected candidates rises more slowly than
$1/Q^4$, as well as more slowly than the background.  Such acceptance
effects may be the reason for the absence of a clear signal at low
$Q^2$ and high $W$. However, this suppression of the signal may also
be related to the unknown production mechanism of the signal.

The invariant-mass spectrum was investigated for the $\ksp$ and
$\kspb$ samples separately. The result is shown as an inset in
Fig.~\ref{pqbar} for $Q^2 > 20 \gev^2$, compared to the fit to the
combined sample scaled by a factor of 0.5. The results for two decay
channels are compatible, though the number of $\kspb$ candidates is
systematically lower.  The mass
distributions were fitted using the same function as the combined
sample and gave statistically consistent results for the peak position
and width (not shown). The number of events in the $\kspb$ channel is
$96 \pm 34$. If the signal corresponds to the $\Theta^+$, this
provides the first evidence for its antiparticle.

If an isotensor state is responsible for the signal, a $\Theta^{++}$
signal might be expected in the $K^+ p$ spectrum\cite{plb570:185}. The
$K^\pm p (K^\pm\bar{p})$ invariant mass spectra were investigated for
a wide range of minimum $Q^2$ values, identifying proton and charged
kaon candidates using $dE/dx$ in a kinematic region similar to that
used in the $\ksppb$ analysis. No peak was observed in the $K^+ p$
spectrum, while a clean $10\sigma$ signal\footnote{
The signal for $\bar{\Lambda}(1520)$ in the $K^+\bar{p}$ spectrum
has the same number of events and significance
as the $\Lambda(1520)$ signal in the $K^-p$ spectrum.}
was observed in the $K^- p$
spectrum at $1518.5\pm 0.6(\mathrm{stat.})\mev$, corresponding to the
$\Lambda(1520)D_{03}$. Performing a fit using a Gaussian fixed to the
detector resolution convoluted with a Breit-Wigner gives an intrinsic
width of $13.7\pm 2.1 (\mathrm{stat.})
\mev$, consistent with the PDG value of $15.6\pm 1.0 \mev$\cite{pr:d66:010001}.

%%%%%%%%%%%%%%%%%%%%%%%%%%%%%%%%%%%%%%%%%%%%%%
\subsection{Systematic checks}
%%%%%%%%%%%%%%%%%%%%%%%%%%%%%%%%%%%%%%%%%%%%%%

A number of checks have been carried out to study possible
reflections from known states and to verify the robustness of the
$1522\mev$ peak.

Tracks from the proton band within twice the width of the $K^*$ peak
were removed. According to a MC calculation,
this cut increased the purity of the proton sample by $15\%$ and
reduced the statistics by a factor of 2.5. The resulting peak position
was unchanged.

The energy of the proton candidates was required to be higher than
that of the $\ks$, to reduce the combinatorial background
\cite{hep-ph-0401122}. Using this cut and the $K^*$-rejection cut, 
a peak may be seen in the mass spectrum even in the $Q^2 > 1
\gev^2$ sample. However, this combination of cuts 
leads to a complicated background shape, making the significance and the
mass of the signal difficult to evaluate.

The $\ks\>K^{\pm}$ invariant mass was reconstructed from the data
using the $K^{\pm}$ mass hypothesis for proton candidates, to see
whether $D_S$ decays contribute to the signal. It was verified that
heavy-flavour particles cannot contribute to this spectrum if $M(\ks
p(\bar{p}) )\le 2000\mev$.

The $\ks$ candidates were combined with primary tracks in the region
$dE/dx\le 1.15$~mips and $p\le 1.5\gev$, where pions are expected to
dominate over (anti)protons. The invariant mass was reconstructed
using the same procedure as before, applying the proton-mass
hypothesis for selected tracks. In addition, the mass distribution was
calculated using the pion band of the $dE/dx$ plot to select proton
candidates, and by using proton and $K^0_S$ candidates from different
events. In none of these cases was any structure seen in the mass
distribution.

The robustness of the peak was also checked by varying the event and
track selections.  The maximum momentum for the proton candidates was
changed in a wide range from $1.2\gev$ to $4.0\gev$.  The $dE/dx$ cut
was varied in the range $1.1$ to $1.3$ mips. No change was seen in the
peak position.

%One could speculate that the observed state is isotensor [S.Capstick
%et al, Phys.Lett. B570 (2003) 185.]  
%%%%%%%%%%%%%%%%%%%%%%%%%%%%%%%%%%%%%%
\subsection{Systematic uncertainties}
%%%%%%%%%%%%%%%%%%%%%%%%%%%%%%%%%%%%%%%

The systematic uncertainties on the peak position and the width,
determined from the fit shown in Fig.~\ref{pqbar}, were evaluated by
changing the selection cuts and the fitting procedure.  The largest
uncertainties on the peak position and the Gaussian width for each check are
given in parentheses (in MeV).  The following systematic studies have been
evaluated:

\begin{itemize}
\item[$\bullet$]
the DIS selection cuts were found to have a negligible effect on the
peak position. The largest uncertainty was found by raising the $Q^2$
cut to $30\gev^2$ ($+0.8$, $+0.3$);

\item[$\bullet$]
the $40$-hit requirement for the proton candidates was not used
($+0.1$, $+1.5$).  The variations of the cuts on the tracks in the
laboratory frame were found to have a negligible effect on the peak;

\item[$\bullet$]
the $dE/dx$ cut was increased to $1.3$~mips ($+0.6$, $-0.7$). The maximum
momentum of protons was varied within $1.3-4.0\gev$
($^{-0.4}_{-0.5}$, $^{+0.8}_{-0.3}$);

\item[$\bullet$]
the bin size was raised and lowered by $1\mev$
($^{-1.2}_{-0.4}$, $^{+0.1}_{+0.6}$ ); the log-likelihood method was used for the
fitting procedure instead of the $\chi^2$ method (no change);

\item[$\bullet$]
the fit was made using only a single Gaussian ($+0.6$,
$-1.2$), or with three Gaussians ($-0.1$, $+0.7$); the
background function was changed to a third-order polynomial
($+0.6$, $0.0$);

\item[$\bullet$]
the CTD momentum calibration uncertainty on the peak position was
calculated from the mass measurements of $\ks$, $\Lambda$ and $K^*$,
as well as the $\Lambda_c$ reconstructed in the $\ksppb$ decay channel
($^{+2.4}_{-0.8}$).

\end{itemize}

The overall systematic uncertainties of $^{+2.8}_{-1.7} \mev$  and $^{+2.0}_{-1.4} \mev$  on the
peak position and the width were 
determined by adding the above uncertainties in quadrature.

%%%%%%%%%%%%%%%%%%%%%%%%%%%%%%%%%%%%%%
\section{Summary and conclusions}
%%%%%%%%%%%%%%%%%%%%%%%%%%%%%%%%%%%%%%

The $\ksppb$ invariant mass spectrum has been studied in inclusive
deep inelastic $ep$ scattering for a large range in the photon
virtuality. For $Q^2 \gtrsim 10\gev^2$ a peak is seen around 1520 MeV.

The peak position, determined from a fit to the mass distribution in
the kinematic region $Q^2\ge 20\gev^2$, is $1521.5\pm 1.5({\rm
stat.})^{+2.8}_{-1.7} ({\rm syst.}) \mev$, and the measured Gaussian
width of $\sigma=6.1\pm 1.6({\rm stat.})^{+2.0}_{-1.4}({\rm
syst.})\mev$ is above, but consistent with, the experimental
resolution of $2.0\pm 0.5\mev$. The number of events ascribed to the
signal by this fit is $221 \pm 48$.  The statistical significance,
estimated from the number of events assigned to the signal by the fit,
varies between $3.9\sigma$ and $4.6\sigma$ depending upon the
treatment of the background. The probability of a similar signal
anywhere in the range 1500--1560 MeV arising from fluctuations of the
background is below $6 \times 10^{-5}$.

The results provide further evidence for the existence of a narrow
baryon resonance consistent with the predicted $\Theta^+$ pentaquark
state with a mass close to $1530\mev$ and a width of less than $15\mev$.
Evidence for such a state has been seen by other experiments, although
the mass reported here lies somewhat below the average mass of these
previous measurements.  In the $\Theta^+$ interpretation, the signal
observed in the $\kspb$ channel corresponds to first evidence for an
antipentaquark with a quark content of
$\bar{u}\bar{u}\bar{d}\bar{d}s$.  The results, obtained at high
energies, constitute first evidence for the production of such a state
in a kinematic region where hadron production is dominated by
fragmentation. 

\section*{Acknowledgements}
\vspace{0.3cm}
We thank the DESY Directorate for their strong support and encouragement.
The remarkable achievements of the HERA machine group were essential for
the successful completion of this work and are greatly appreciated. We
are grateful for the support of the DESY computing and network services.
The design, construction and installation of the ZEUS detector have been
made possible owing to the ingenuity and effort of many people from DESY
and home institutes who are not listed as authors.

\vfill\eject

%% file: DESY-04-056-ref.tex
{
\def\bibname{\Large\bf References}
\def\refname{\Large\bf References}
\pagestyle{plain}
\ifzeusbst
  \bibliographystyle{./BiBTeX/bst/l4z_default}
\fi
\ifzdrftbst
  \bibliographystyle{./BiBTeX/bst/l4z_draft}
\fi
\ifzbstepj
  \bibliographystyle{./BiBTeX/bst/l4z_epj}
\fi
\ifzbstnp
  \bibliographystyle{./BiBTeX/bst/l4z_np}
\fi
\ifzbstpl
  \bibliographystyle{./BiBTeX/bst/l4z_pl}
\fi
{\raggedright
\bibliography{./BiBTeX/user/syn.bib,%
              ./BiBTeX/bib/l4z_articles.bib,%
              ./BiBTeX/bib/l4z_books.bib,%
              ./BiBTeX/bib/l4z_conferences.bib,%
              ./BiBTeX/bib/l4z_h1.bib,%
              ./BiBTeX/bib/l4z_misc.bib,%
              ./BiBTeX/bib/l4z_old.bib,%
              ./BiBTeX/bib/l4z_preprints.bib,%
              ./BiBTeX/bib/l4z_replaced.bib,%
              ./BiBTeX/bib/l4z_temporary.bib,%
              ./BiBTeX/bib/l4z_zeus.bib}}
}
\vfill\eject

%% file: DESY-04-056-tab.tex
%-------------------------------------------------------------------------------
%       table
%------------------------------------------------------------------------------
\begin{table}
\centering
\begin{tabular}{|c|c|c|c|c|c|c|c|} \hline
Fit
&
&
Gaussian+Bkg.
&2 Gaussians + Bkg.\\
\hline
$\chi^2/\mathrm{ndf}$
&$M\leq 1700$ MeV
&51/47
&35/44\\
\hline
&mass (MeV)
&-
&$1465.1\pm 2.9$\\
Peak 1
&width (MeV)
&-
& $15.5\pm 3.4$\\
&events
&-
& $368\pm 121$\\
\hline
&mass (MeV)
&$1522.2\pm 1.5$
& $1521.5\pm 1.5$ \\
Peak 2
&width (MeV)
&$4.9\pm 1.3$
& $6.1 \pm 1.6$\\
&events
&$155\pm 40$
&$221\pm 48$\\
\hline
\end{tabular}
\caption{
Fit results for $Q^2>20\gev^2$.
}
\label{table1}
\end{table}

\vfill\eject

%% file: DESY-04-056-fig.tex
%-------------------------------------------------------------------------------
%       Results
%-------------------------------------------------------------------------------

\begin{figure}
\begin{center}
  \includegraphics[height=7.0cm]{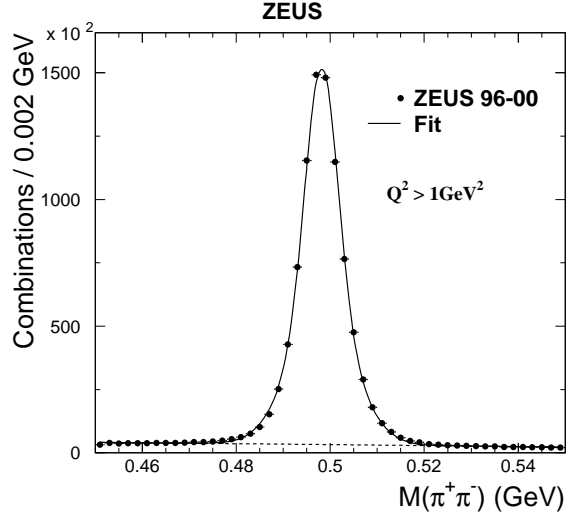}%
\caption{
The $\pi^+\pi^-$ invariant-mass distribution for $Q^2>1\gev^2$. 
The solid line shows the fit result using a double Gaussian plus a
linear background, while the dashed line shows the linear background.
}
\label{pap0}
\end{center}
\end{figure}

%%%%%%%%%%%%%%%%%%%%%%%%%%%%%%%%%%%%%%%%%%%%%%%%%%%%
\begin{figure}
\begin{center}
\begin{minipage}[c]{0.5\textwidth}
  \includegraphics[height=7.0cm]{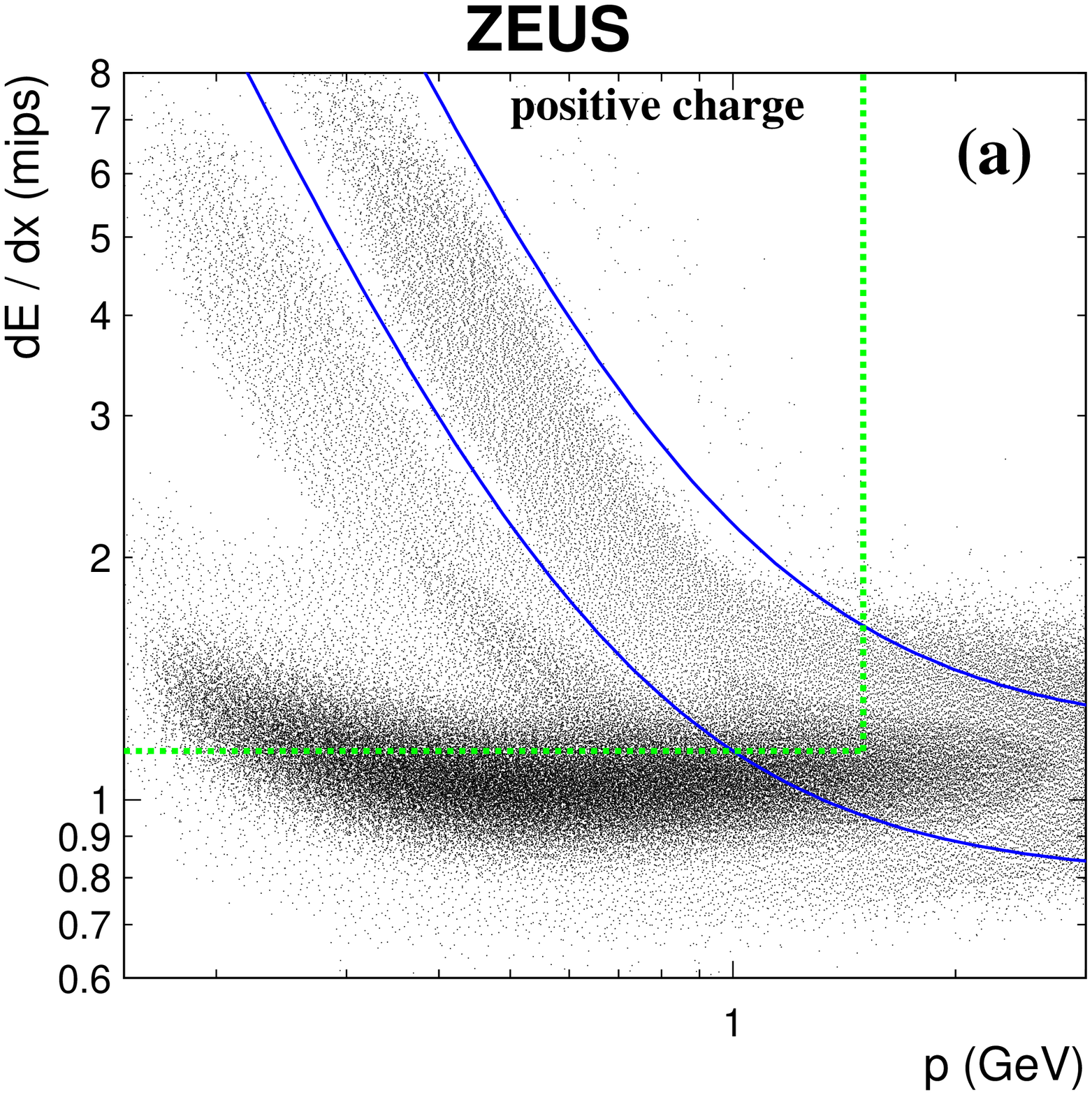}
\end{minipage}%
\begin{minipage}[c]{0.5\textwidth}
  \includegraphics[height=7.0cm]{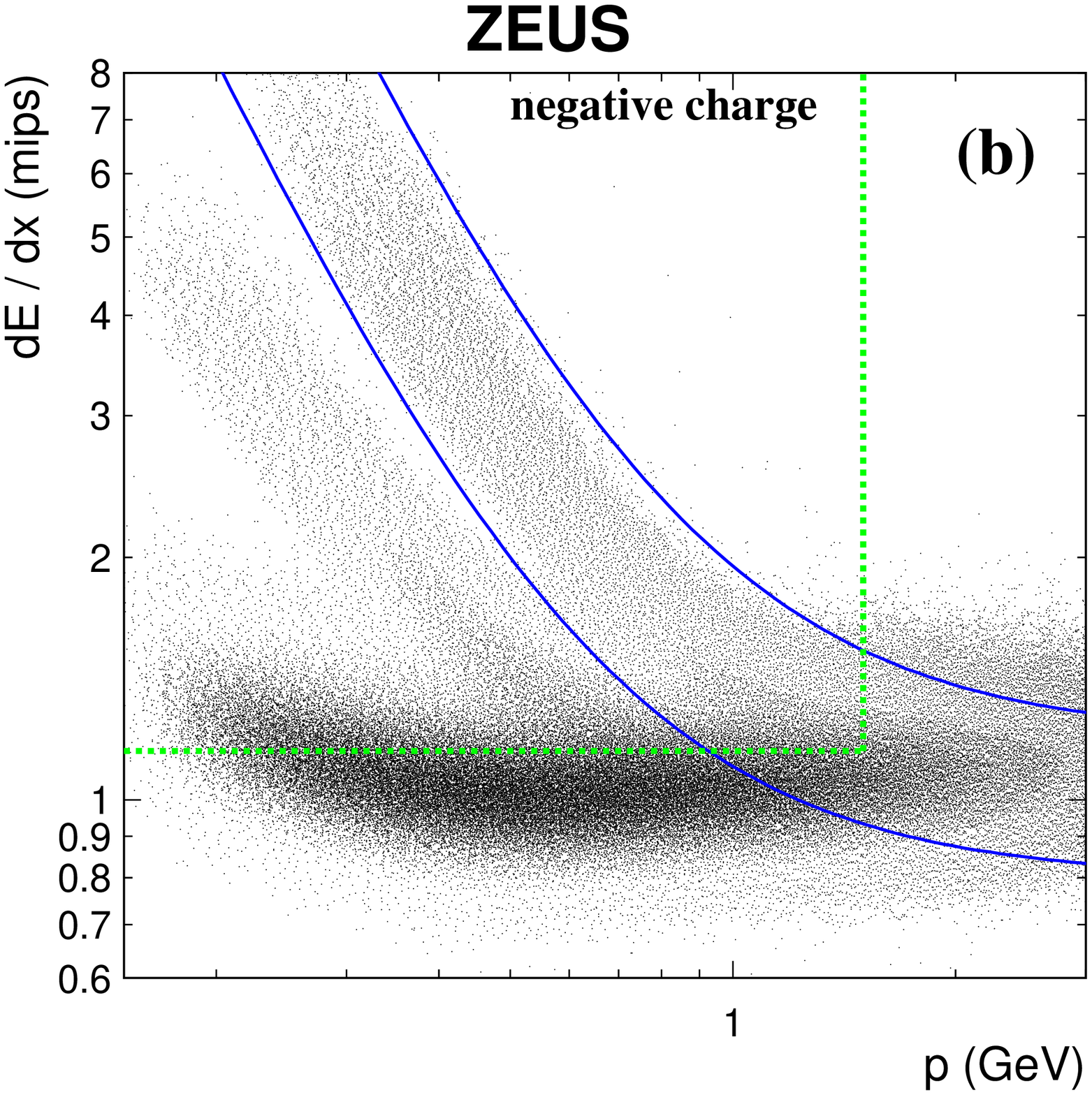}
\end{minipage}
\caption{ The $dE/dx$ distribution as a function of the track momentum
for: {\bf (a)} positive and {\bf (b)} negative tracks. The $dE/dx$ is
normalised to a minimum ionising particle, defined as the average
truncated mean of pion tracks in the momentum range
$0.3<p<0.4\gev$. The solid lines indicate the (anti)proton bands used
in this analysis, and the dotted line denotes the cuts $dE/dx > 1.15$~mips
and $p < 1.5\gev$.}
\label{pap0dedx}
\end{center}
\end{figure}

\begin{figure}
\begin{center}
\includegraphics[height=17.1cm]{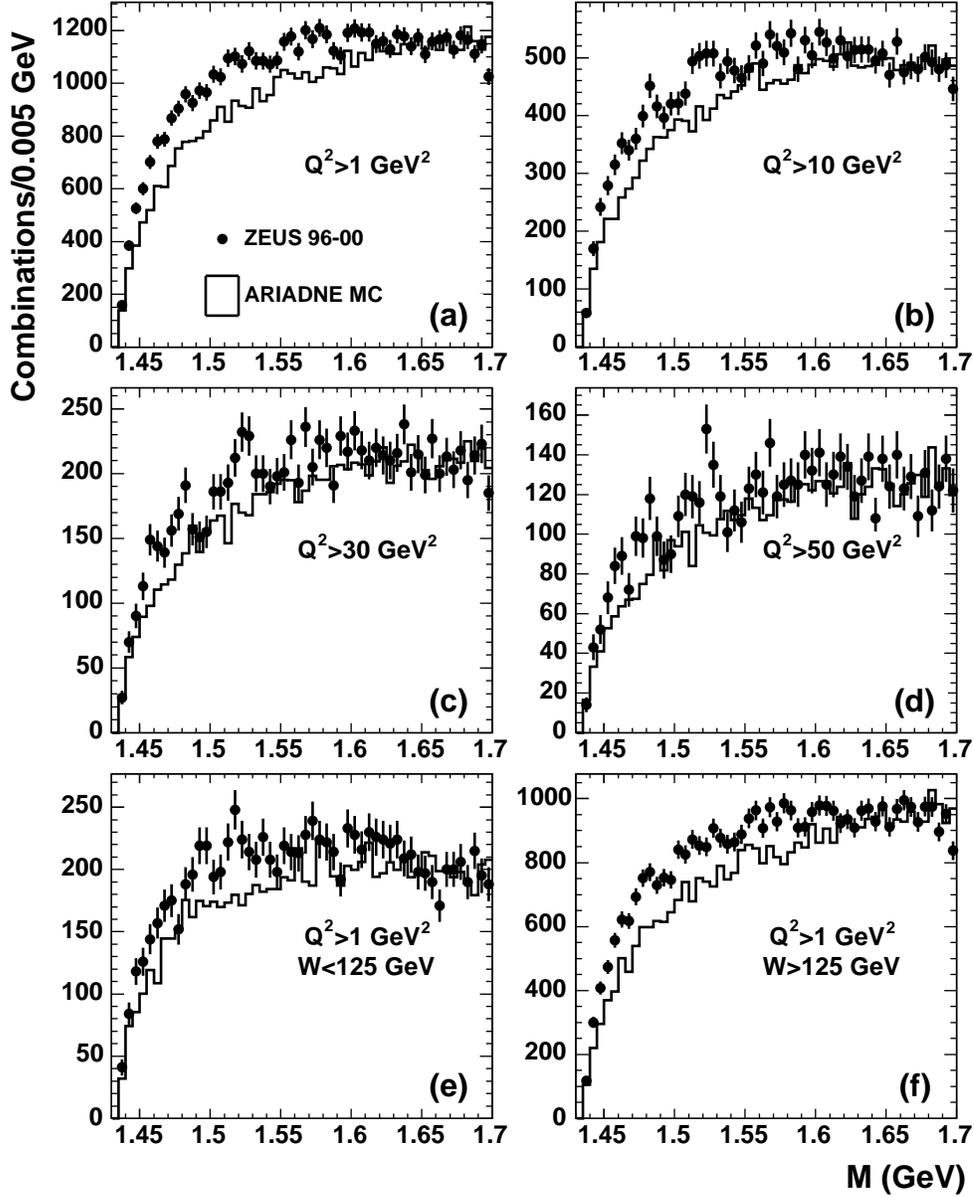}
\caption{ Invariant-mass spectrum for the $\ksppb$ channel, after the
cuts $f\le dE/dx \le F$, $dE/dx>1.15$~mips and $p<1.5\gev$, integrated for
(a) $Q^2 > 1 \gev^2$, (b) $Q^2 > 10 \gev^2$, (c) $Q^2 > 30 \gev^2$, (d)
$Q^2 > 50 \gev^2$, (e) $Q^2 > 1 \gev^2$ and $W<125 \gev$ (f) $Q^2 > 1
\gev^2$ and $W>125 \gev$. The histogram shows the prediction of the
{\sc Ariadne} MC simulation normalised to the data in the mass region
above $1650\mev$.  }
\label{pq}
\end{center}
\end{figure}

%%%%%%%%%%%%%%%%%%%%%%%%%%%%%%%%%%%%%%%%%%%%%%%%%%%%
\begin{figure}
\begin{center}
\includegraphics[height=18.1cm]{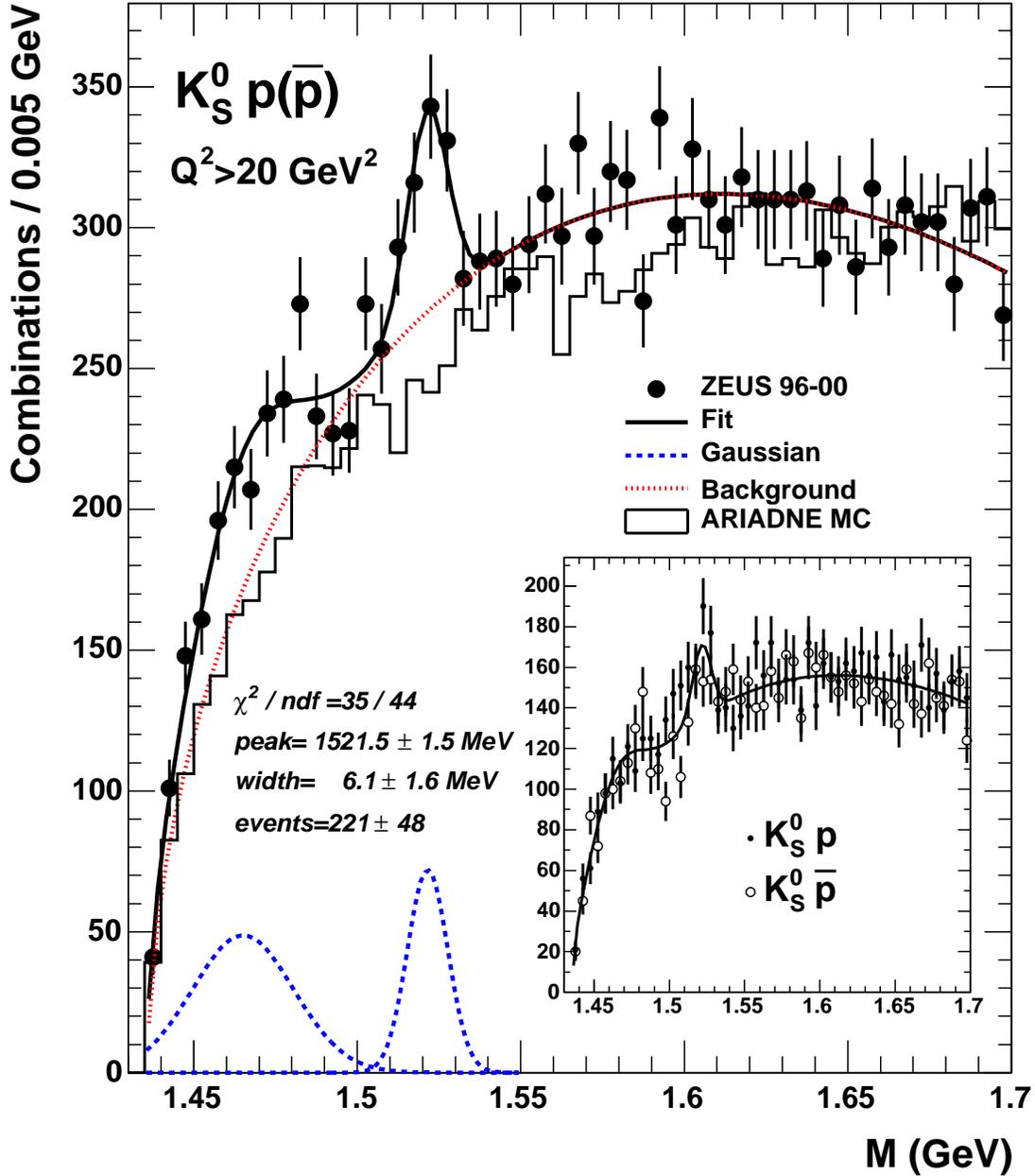}
\caption{ Invariant-mass spectrum for the $\ksppb$ channel for $Q^2 >
20 \gev^2$, with other cuts as in Fig.~\ref{pq}. The solid line is the
result of a fit to the data using a three-parameter background
function plus two Gaussians (see text).  The dashed lines show the
Gaussian components and the dotted line the background according to
this fit. The histogram shows the prediction of the {\sc Ariadne} MC
simulation normalised to the data in the mass region above
$1650\mev$. The inset shows the $\kspb$ (open circles) and the $\ksp$
(black dots) candidates separately, compared to the result of the fit
to the combined sample scaled by a factor of 0.5.  }
\label{pqbar}
\end{center}
\end{figure}